\documentclass[useAMS,usenatbib,usegraphicx]{mn2e}
\usepackage{times, subfigure}
\newcommand{\degree}{\ensuremath{^\circ}}

\title[SEDs of type 2 QSOs]{Spectral Energy Distributions of type 2 QSOs: 
obscured star formation at high redshifts}
\author[D.~Rigopoulou et al.]{D.~Rigopoulou,$^{1}$\thanks{E-mail:
d.rigopoulou1@physics.ox.ac.uk} V.~Mainieri,$^{2,8}$ O.~Almaini,$^{3}$ 
 A.~Alonso-Herrero,$^{4}$ J-S.~Huang,$^{5}$
\newauthor  G.~Hasinger,$^{2}$ G.~Rieke,$^{6}$,
J.~Dunlop$^{7}$ and I.~Lehmann$^2$\\
$^{1}$Department of Astrophysics, Oxford University, Keble Road, Oxford, 
OX1 3RH, UK\\
$^{2}$Max-Planck-Institut fuer extraterrestrische Physik (MPE), 
Postfach 1312, Garching 85741, Germany\\
$^{3}$School of Physics and Astronomy, University of Nottingham, University 
Park, Nottingham NG7 2RD\\
$^{4}$Instituto 
de Estructura de la Materia, CSIC, E-28006 Madrid, Spain\\
$^{5}$Harvard-Smithsonian Center for Astrophysics, 60 Garden Street,
Cambridge, MA 02138, USA\\
$^{6}$Steward Observatory, University of Arizona, 933 North Cherry Avenue, 
Tucson, AZ 85721, USA\\
$^{7}$Institute for Astronomy, University of Edinburgh, Blackford Hill, 
Edinburgh EH9 3HJ, UK\\
$^{8}$European Southern Observatory, Karl-Schwarschild-Strasse 2, D--85748 
Garching b. Muenchen, Germany\\
}
\begin{document}

\date{}

\pagerange{\pageref{firstpage}--\pageref{lastpage}} \pubyear{2007}

\maketitle

\label{firstpage}

\begin{abstract}

We present new mid-infrared and submillimetre observations for a sample of 
eight high redshift type-2 QSOs located in the Chandra Deep Field South. The 
sources are X-ray absorbed with luminosities in excess of 
10$^{44}$ erg s$^{-1}$. Two of the targets have robust detections, 
$S/N >$ 4, while a further three targets are marginally detected with
$S/N \geq$2.5.
All sources are detected in multiple mid-infrared bands with the Spitzer
Space Telescope. 
The multiwavelength spectral energy distributions (SEDs) 
of the type-2 QSOs are compared to those of two local ultraluminous galaxies
(Arp220 and IR22491) in order to assess contributions from a star-forming
component in various parts of the SED. We suggest that their submillimetre 
emission is possibly due to a starburst while a large fraction of the 
mid-infrared 
energy is likely to originate in the obscured central quasar. 
Using the mid-infrared and submm observations we derive infrared 
luminosities which are found to be in excess of L$>$10$^{12}$L$_{\odot}$.
The submillimetre (850 $\mu$m) to X-ray (2 keV)
spectral indices ($\alpha_{SX}$) span a wide range. About half of 
the type-2 
QSOs have values typical for a Compton-thick AGN with only 1 per cent of the 
nuclear emission seen through scattering and, the remaining with values typical
of submm-bright galaxies. Combining the available observational evidence we 
outline a possible scenario for the early stages of evolution of these sources.

\end{abstract}


\section{Introduction}

The discovery that most local spheroids contain massive black holes
(MBH) in  their centres has renewed the interest in the physical
properties of active galactic nuclei. A number of correlations, such
as the one between the masses of MBH and the properties of the host
galaxies (e.g. Gebhardt et al. 2001, Merritt \& Ferrarese 2001) and the 
proportionality
between the mass and the velocity dispersion of the stars, 
suggest a direct link between the
formation$/$growth of the black hole and the stellar mass of the
galaxy spheroid.

The availability of major facilities in the submm (e.g. SCUBA on the
James Clerk Maxwell Telescope and MAMBO on IRAM) led to the discovery
of a large population of luminous infrared galaxies at high redshifts
(e.g. Hughes et al. 1998, Bertoldi et al. 2000, Borys et al. 2003).  
Followup observations in CO molecular
gas-line emission suggest that such galaxies are in fact massive
(e.g. Greve et al. 2005, Chapman et al. 2008).  The launch of the
Spitzer Space Telescope (SST, Werner et al. 2004) has shed new light into
luminous dusty high redshift galaxies. Sensitive mid-infrared
observations have unveiled the presence of obscured AGN whose properties 
would have otherwise remained unnoticed (e.g. Lacy et al. 2004, 
Stern et al. 2005, Alonso-Herrero et al. 2006 hereafter AH06, Donley et 
al. 2007). 

There are two obvious ways one can investigate the coeval formation of
black holes and galactic bulges. One approach is to follow up on known
mm$/$submm sources with X-rays. Somewhat surprisingly, however, only a
small fraction of SMGs contain a visible luminous QSO (e.g. Fabian et al. 1999,
Almaini et al. 2003). Such a result implies that the epoch of black
hole formation does not coincide with the epoch of major star-forming
activity. Using deep X-ray data Alexander et al. (2005) 
found that $\sim$75\% of the Chapman et al. (2005) SCUBA sources do contain 
an AGN however, the bolometric output is dominated by intense star formation.
This claim has more recently been confirmed by mid-infrared studies of
submm-luminous sources by Pope et al. (2008). 
The second approach is to select ``obscured AGN'' and follow
them up in the submillimetre wavelengths. Such an approach has already
been adopted by Page et al. (2001), Archibald et al. (2001) and, Stevens 
et al. (2005) with encouraging results. These authors selected targets
from the ROSAT surveys (e.g. Page et al. 2001), XMM-Newton and Chandra 
(e.g. Page et al. 2003) that span the redshift range 1$<$z$<$3, luminosity
range 44.1$<$log L$_{X}<$44.4 and absorbing column densities
21$<$log N$_{H}$$<$23. 

The advantage of the present work is that we are
using a sample of ``heavily'' X-ray absorbed (log N$_{H} >$ 23
cm$^{-2}$) and very luminous (log L$_{X} >$ 44 erg s$^{-1}$) AGN, the so
called type-2 QSOs, located in the Chandra Deep Field South (CDF-S) and
selected from Szokoly et al. (2004).  The first submm detection of
such a heavily absorbed QSO has been reported in Mainieri et
al. (2005).

In this paper we combine the submm observations of the entire sample
of eight type-2 QSOs with observations obtained with SST. We discuss the
spectral energy distribution of the sources and compare them with
local templates. We investigate a possible link between type-2 QSOs
and SMGs, especially those known to contain a black hole. We explore
the origin of the mid-infrared and submillimeter emission and finally
discuss a possible evolutionary scheme for type-2 QSOs.
Throughout the paper we assume a cosmology $\Omega_{M}=0.3$, 
$\Omega_{\Lambda}=0.7$ and H$_{0}$ = 70 km s$^{-1}$ Mpc $^{-1}$.

\section{Sample Selection and Observations}

\subsection{The type-2 QSO Sample}

The deep ($\sim$1 Msec) Chandra survey of the Chandra Deep Field South (CDF-S,
Giacconi et al. 2002) resulted in the discovery of a unique sample of heavily
X-ray absorbed (log N$_{H}$ $>$ 22 cm$^{-2}$) and very luminous 
(log $Lx >$ 44 erg s$^{-1}$) AGN, the so-called type-2 QSOs. 
The type-2 QSOs for this program have been selected from the deep 
($\sim$ 1 Msec) survey of the CDF-S (Szokoly et al. 2004). 
From the 10 objects contained in the original list, 7 objects were observed 
with the JCMT within the allocated time. To this sample, source XID 901, was 
added. Although the source does not meet the formal $L_{x}$ criterion 
dscribed above, it is a heavily absorbed object at $z>2$. In Table 1 we 
list properties of the sample of 
type-2 QSOs studied here: names, coordinates, redshifts, X-ray luminosities 
and absorbing column densities. We note that X-ray luminosities are corrected
for absorption (values taken from Tozzi et al. 2006). 

\begin{table*}
 \centering
 \begin{minipage}{180mm}
  \caption{Optical coordinates, spectroscopic redshifts, X-ray luminosities and 
absorbing column densities}
  \begin{tabular}{@{}ccccccc@{}}
  \hline
ID$^{1}$&RA(2000)&Dec(2000)&L$_{X}[0.5-2 KeV] ^{2,3}$&L$_{X}[2-10 KeV]^{2,3}$ &N$_{H} /$ 10$^{22}$ $^{2}$&z\\
        &  &  & 10$^{43}$erg s$^{-1}$&10$^{44}$erg s$^{-1}$  &(cm$^{-2}$)  &  \\
 \hline
202&  03 32 29.86&   -27 51 05.8&19.8&5.69 &150.0  &     3.700\\ 
54 &  03 32 14.61&   -27 54 20.7&3.03&0.88 &10.67$^{+5.40}_{-4.57}$  & 2.561  \\
45 &  03 32 25.68&   -27 43 05.7&4.03&1.10 &8.19$^{+3.02}_{-2.66}$   &     2.291\\
263&  03 32 18.83&   -27 51 35.6&8.67&2.91   & 150.0     &     3.660 \\
27 & 03 32 39.67&   -27 48 50.5&5.23&2.13     &28.08$^{+9.18}_{-7.97}$      &     3.064 \\
112&  03 31 52.07&   -27 53 28.2&7.52&1.16 &28.99$^{+8.89}_{-4.86}$      &     2.940 \\
901&  03 32 35.78&   -27 49 16.82&0.87&0.13 &18.94$^{+17.86}_{18.15}$     &     2.578\\
51 &  03 32 17.16 &  -27 52 20.7&5.88&1.02 &22.42$^{+2.85}_{-2.44}$      &     1.099\\ 
\hline

$^{1}$ : IDs are from Giacconi et al. (2002)\\
$^{2}$ : N$_{H}$ values from Tozzi et al. (2006)\\
$^{3}$ : X-ray luminosities corrected for absorption.\\
\end{tabular}

\end{minipage}
\end{table*}

\subsection{Submillimeter data}

Observations at 850 $\mu$m were carried out on the JCMT during 2004
August and November. We used the Submillimetre Common User  Bolometer
Array (SCUBA, Holland et al. 1999) in photometry mode, in which the
source is placed on the central bolometer of the array and the
secondary mirror is jiggled in a 3$\times$3 pattern with 2 arcsec
intervals with a 1 sec integration at each position. The secondary mirror
was chopped 45 arcsec in azimuth at a frequency of 7.8 Hz and nodded
between the source and reference positions every 18s. We placed each
source in the central bolometer (H7) and used the median  of the
remaining bolometers for additional sky removal. We used Uranus to
calibrate the derived flux densities. Calibration uncertainties are
about 10\% at 850 $\mu$m.  The pointing of the telescope was checked
frequently while the sky opacity was monitored via regular skydips using
the JCMT Water Vapour Monitor and the CSO (Caltech Submillimetre Observatory)
Tau Meter. Observations
during August 2004 were made under very good weather conditions with the
225- GHz sky opacity as measured at the adjacent CSO, $\tau_{225}$, 
in the range 0.05--0.13 (at the airmass of the target).

We reduced data using the standard STARLINK software collection SURF.
After compensating for the nod the data were flatfielded and corrected
for atmospheric extinction. Each jiggle in turn was then corrected for
residual sky noise which is correlated across the SCUBA field of view
and often dominates the signal from faint sources. After
sky-subtraction the data were clipped at the 3-$\sigma$ level. Since
data for the same source were collected on different nights we tested
them for consistency with one another using a Kolmogorov-Smirnov test,
rejecting anything below the 5\% mark. Submm fluxes are reported in Table 2.
Two sources were detected at $S/N \geq$ 4, while a further 3 are observed
at $S/N\geq$ 2.5 level. We note that CDFS-202 is the only type-2 QSO for
which the SCUBA observations yielded no signal at all and therefore a
formal upper limit (corresponding to the sensitivity limit achieved by
the observations) is quoted.  
%Therefore, this object has been omitted
%in subsequent %analysis.

\subsection{Spitzer Observations}

The CDF-S was observed 
at 24 and 70 $\mu$m
with the Multiband Imaging Photometer for Spitzer (MIPS, Rieke et al. 2004) 
and the InfraRed Array Camera (IRAC, Fazio et al. 2004) 
covering a total area of $1^{\degree}.5 \times$0$^{\degree}.5$ as part
of the MIPS Guaranteed Time Observations (GTO). Additional observations
of the central region of CDF-S 
were taken as part of the Great Observatories Origins Deep Surveys 
(GOODS, for an overview see Dickinson et al. 2003).
Finally, additional 70 $\mu$m data were taken as part of 
the FIDEL (P.I. M. Dickinson) program. 

The 24$\mu$m source  extraction and photometry is described in detail in
e.g. Papovich et al. (2007).  In summary, all sources were treated as
point-like, given the $\sim$5$^{''}$.8 FWHM angular resolution. PSF
fitting and flux measurements were performed using  DAOPHOT
packages. The resulting 5 $\sigma$ flux density limit was 80 $\mu$Jy
reaching a completion of 75\% at this level.
IRAC observations were carried out at 3.6, 4.5, 5.8 and 8.0 $\mu$m.
Source detection, extraction and 
photometry was performed in a similar manner to the MIPS data i.e. using 
DAOPHOT routines. IRAC photometry was carried out 
using a PSF of 1$\farcs8$--2$\farcs0$. The aperture fluxes in each band were 
subsequently corrected to total fluxes using known PSF growth curves 
from Fazio et al. 2004; Huang et al. 2004. The resulting 5$\sigma$
flux density limits were 1.73, 3.02, 10.96 and 8.3 $\mu$Jy at 3.6, 4.5, 5.8
and 8.0 $\mu$m, respectively.

The Spitzer photometric
data are listed in Table 2. Only one (XID 901) of the sources was detected
at 70$\mu$m (B. Weiner, prov. comm.).

\begin{table*}
 \centering
\begin{minipage}{140mm}
  \caption{Mid-Infrared and submm fluxes and 1$\sigma$ uncertainties}
  \begin{tabular}{@{}cccccccc@{}}
  \hline
ID &3.6&4.5&5.8&8.0&24&70&850\\
  &($\mu$Jy) &($\mu$Jy)&($\mu$Jy)&($\mu$Jy)&($\mu$Jy) &($\mu$Jy) &(mJy) \\
\hline
202&3.82(0.5)&1.63:$^{1}$&4.69(0.5)&10.47(0.5) &76.10(0.5)&0.0&2.0(2.6)\\
54&4.4(0.5)&0.0$^{2}$&10.1(1.4)&5.4$^{1}$&60(9)&0.0&9.74(3.65)\\
45&14.6(2.2)&21.8(3.5)&53.8(5.0)&124.6(23)&480(39)&0.0&2.22(1.09)\\
263&7.68(0.6)&6.82(1.1)&11.3(1.9)&45(3.5)&63(8)&0.0&4.82(1.11)\\
27&13.6(1.2)&15.5(2.2)&18.9(3.4)&24.6(3.0)&154(22)&0.0&4.81(3.45)\\
112&6.4(0.7)&9.6(1.0)&18.1(2.7)&34.7(3.7)&301(49)&0.0&9.87(1.76)\\
901&14.9(1.5)&16.7(1.9)&30.3(3.5)&41.0(3.9)&520(45)&3320(1850)&8.8(3.6)\\
51&66(5.7)&73(6.5)&0.0&99(22)&0.0&0.0&2.89(1.06)\\
\hline

$^{1}$ : denotes a (3$\sigma$) upper limit\\
$^{2}$: source completely undetected.

\end{tabular}
\end{minipage}
 
\end{table*}

\section{Spectral Energy Distributions}

In Figure 1 we show rest-frame (Spectral Energy Distributions (SEDs) for the 
present sample of type-2 QSOs. 
\begin{figure*}
\begin{minipage}{180mm}
 \centering
  \includegraphics[width=140mm, angle=270]{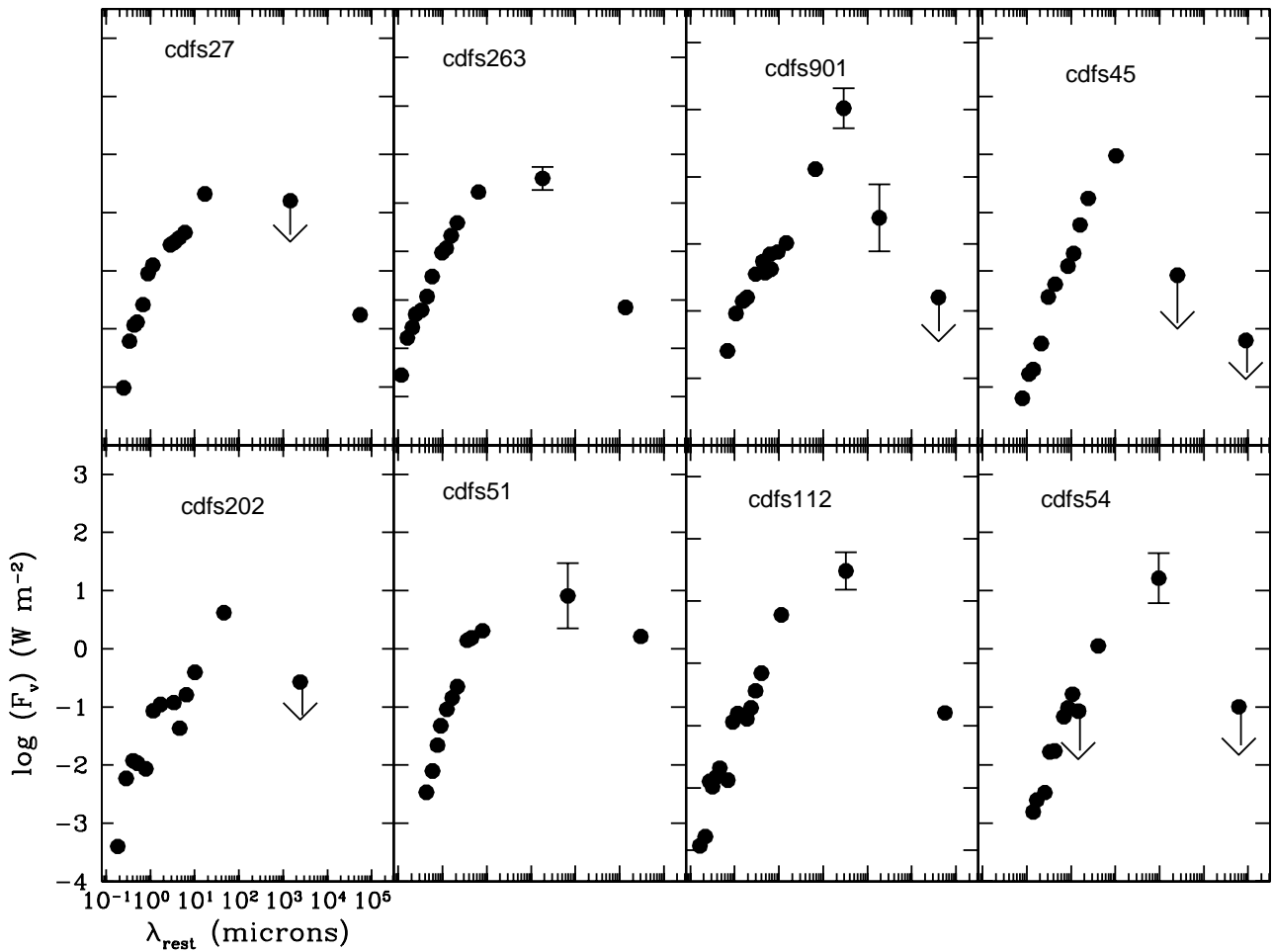}
\caption{Spectral Energy Distributions of type-2 QSOs. The photometric
points include  optical HST/ACS data (Giavalisco et al. (2004), near
infrared data (Rettzlaff et al. 2009), Spitzer and SCUBA data
presented here (shown in Table 2), X-ray detections from Giacconi et
al. (2002) and radio data from Kellermann et al. (2008) and 
Miller et al. (2008). Errorbars are shown for the submm data and the MIPS
70 $\mu$m data. In the remaining cases the errorbars are smaller than
the plotting symbols (see Table 2).
}
\end{minipage}
\end{figure*}
To construct the rest-frame SEDs we used 
optical photometry from HST$/$ACS observations (as part of the GOODS 
survey, Giavalisco et al. 2004), near-IR ISAAC imaging data 
(Retzlaff et al. 2009), IRAC$/$MIPS Spitzer 
observations, and X-ray data (Giacconi et al. 2002). 
The radio data come from the VLA survey of the CDF-S, Kellerman et al (2008),
Miller et al. (2008). 
Spectroscopic redshifts for all
the type-2 QSOs presented here are reported in Szokoly et al. (2004). 
The overall shape of the SED of the type-2 QSOs is rather different
from the standard (type-1) QSO SED (see discussion for CDFS-263 in Mainieri et 
al. 2005). From the shape of the SEDs presented in Figure 1 and the
discussion in Mainieri et al. (2005) it is clear that
type-2 QSOs and, in that instance CDFS-263, do not show the featureless
steep continuum expected for a typical QSO. Instead, the shape of their SEDs
implies the presence of an additional starburst component. 

Here we adopt a slightly different approach and using templates chosen
from local Ultra-luminous Infrared Galaxies (ULIRGs) we try to assess
the dominant contribution of starburst and$/$or AGN components in the
various parts of the type-2 QSO SEDs. Our choice of templates (ULIRGs) was 
driven by the derived far-infrared luminosities of type-2 QSOs which are
discussed in section 6.
We stress that the aim of this exercise is not
to find out whether there is an AGN in these type-2 QSOs (by
definition they have a central massive accreting black hole), instead,
we want to assess the relative contribution of the the two components (AGN
or starburst) in  various parts of the SED. In particular we focus on
the near-infrared and submm wavelengths; it is clear that the mid-infrared 
is dominated by emission from the AGN. We chose
two templates, the highly obscured ULIRG archetype Arp220 and the more
typical (less obscured) starburst-dominated ULIRG IR22491 so that we can
explore the effect of obscuration in the type-2 QSOs.

In Figure 2 we show the two templates and the SEDs of the type-2 QSOs presented
in this study. The SEDs of the two templates as well as those of the type-2 
QSOs have been normalised in the submm. It is clear that in all cases (all
type 2 QSOs) an extra AGN component is immediately needed to explain
the mid-infrared part of the SED. 
Two of the sources, CDFS-27,
and CDFS-263 are consistent with the SED of IR22491 (ie a
typical ULIRG) with an extra component that comes out in the rest-frame 
6--30 $\mu$m range. The near-infrared points follow the SED of the template
implying that the near-infrared light comes mainly from the stellar population
in the (massive) host galaxy. The SED of CDFS-45 rises more steeply 
than the SED of IR22491, possibly due to a stronger underlying AGN and$/$or 
less obscuration.

Sources CDFS-112, CDFS-901 and CDFS-202 are consistent with the
Arp220 SED plus again an additional AGN component to account for emission in
the mid-infrared. The difference with these sources is that they 
require a more heavily absorbed starburst component to match the 
near-infrared part of the SED.
We caution however, that this finding
does not imply an abundance of Arp 220-like objects at redshift of
$\sim$3, instead, we suggest that these sources are heavily obscured
with rather unusual properties. Rigby et al. (2008) discuss the
the lack of appropriate templates for targets at redshifts
2--3. The remaining source CDFS-54, poses a real challenge when fitting 
the SED as it is undetected at 8 but detected at 24 microns. The SED of 
CDFS-54 falls below the SEDs of both templates considered here. 

 Although the current sample is small, the agreement
 between the templates and the SEDs of the
type-2 QSOs suggests that type-2 QSOs might have 
a significant starburst component.
Star-forming processes could account for both
the visible$/$near-infrared light as well as the submm.
The mid-infrared, however, is in all cases powered by the central AGN. 
We note that the SEDs of the
type-2 QSOs show a greater variety in shape than the SEDs of SCUBA
galaxies (c.f. Pope et al. 2008). In the case of the type-2 QSO
population this may reflect the various stages of evolution of the
galaxies which may also affect the energetics (starburst vs central
AGN). 
\begin{figure*}
\includegraphics[width=140mm, angle=270]{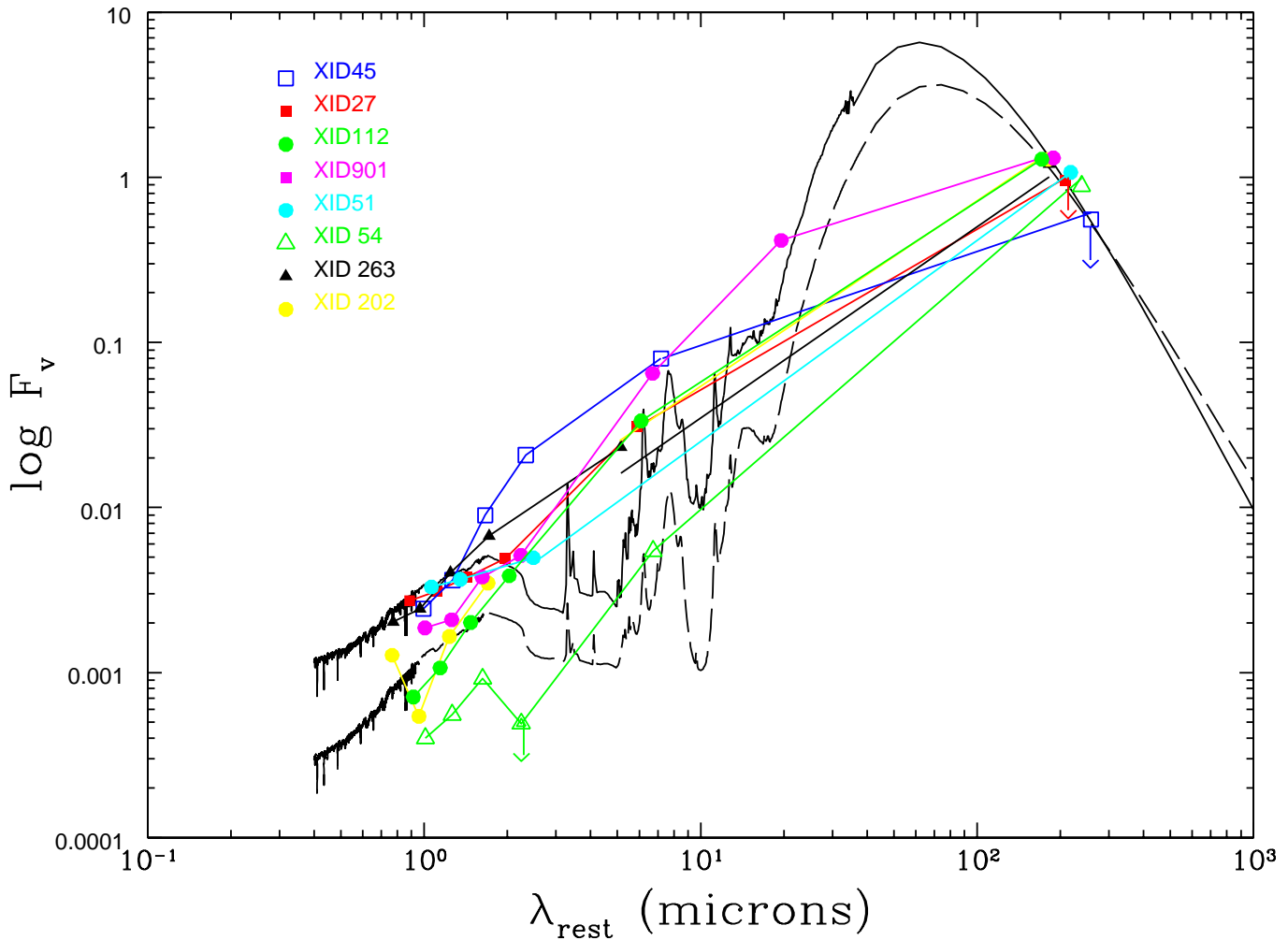}
 \caption{Comparison of the Type 2 QSO SEDs with those of template
galaxies Arp220 (long dashed line) and IR22491(solid line). The template
SEDs have been normalised at 850 $\mu$m. 
The photometry used to construct the SEDs of type 2 QSOs has been described
in Figure 1.}
\end{figure*}

Interestingly, two of the QSOs presented in our study are also part of
the mid-IR selected powerlaw sample of AH06. CDFS-901 and CDFS-112 have 
been classified as
Narrow-Line AGN (NLAGN) and ULIRG-like, respectively. Since, however,
the division between the two  categories is rather arbitrary we will,
as in AH06, consider them as one  class. According to AH06, galaxies
in this category show a steep powerlaw  continuum with a less
prominent 1.6 micron bump. Both CDFS-901 and CDFS-112 follow the Arp 220 
template SED (Figure 2), except for the mid-infrared where the additional
need for a steep power-law (AGN) is evident.
Overall, we find a dichotomy in the shape of the type-2 QSO SEDs
which likely reflects the varying contribution of the starburst component
to the overall energy output of the sources. While starburst-related 
activity is responsible for the energy released in the submillimetre, 
the mid-infrared is likely the place for the dominance of the central compact
source which we discuss next.

\section{Mid-Infrared Continuum}

At the redshift of our type-2 QSOs the IRAC bands probe the rest-frame 
near-infrared continuum while, the MIPS 24$\mu$m 
band probes the mid-infrared around 6$\mu$m which is significantly 
affected by the obscuring dust. The 24$\mu$m flux could originate either
from emission related to star-forming processes (at this redshift the
6.2 $\mu$m Poly-Aromatic Hydrocarbon (PAH) feature enters the MIPS 
24$\mu$m band) 
or it could contain indirect 
contributions from the AGN via heating of the surrounding dust which in turn 
emits in the mid-infrared. Due to the inherited complexity of the 
mid-infrared 
part of the spectrum, we will not attempt a decomposition 
of the mid-infrared part of the SED of type-2 QSOs but instead
opt for simple colour-colour
mid-infrared diagnostic diagrams.

A number of authors have published results on AGN selection
based on mid-infrared colours (e.g Lacy et al. 2004, Stern et al. 2005, 
Hatziminaoglou et al. 2005 but see Barmby et al. 2006, Rigby et al. 2008
for a discussion of the various criteria in detail).
The Lacy et al. (2004) plots involve
IRAC and MIPS bands only and are, thus, more appropriate for our purposes. 
Here, we examine the mid-infrared colours of our sources and compare
them to those of high-z submillimetre luminous galaxies (hereafter SMGs). 
bf Based on the shape of their SEDs,  SMGs can be crudely 
divided in those that display a 
clear 1.6$\mu$m ``stellar bump" and their overall SED shape is similar 
to that of Arp220
(referred to as "cold" SMGs)
and, those with  a steep
powerlaw SED similar to the SED of Mrk231 (referred to as warm, see 
e.g. Egami et al. 2004).
Figures 3a and 3b show the Lacy et al. (2004) mid-infrared
colour-colour criteria for our type-2 QSOs and 
SMGs from Egami et al. (2004) and Ashby et al. (2006). 
In the IRAC plot (Fig. 3a) we find that
CDFS-27, CDFS-45 and CDFS-263 display the reddest colours both in 
$S_{8.0}/S_{4.5}$ and $S_{5.8}/S_{3.6}$ and thus are similar to those of 
"warm" SMGs. These three type 2 QSOs were best fit by the less obscured 
IR22491 template (of course as discussed in Section 3 an AGN component 
is necessary to explain the mid-infrared part of the spectrum). 
Similarly, type 2 QSOs following the Arp220 SED 
show IRAC colours with values closer to those of ``cold'' SMGs. 
In the IRAC$/$MIPS plot (Fig. 3b) the colours of type 2 QSOs vary
over a wider range (especially the 24/5.8 ratio). AGN display colours
that would fall within the shaded area in Fig. 3b. As expected, ``warm''
SMGs fall within the shaded region. Among the type 2 QSOs, those that
were fit with the less-obscured ULIRG template IR22491 (plus AGN) 
fall within the shaded
region expected for AGN. The remaining type 2 QSOs fall outside the expected
range for AGN. 

Although the present type-2 QSO sample is small, the mid-infrared
colour-colour plots allow us to draw some qualitative
conclusions. First, in both plots the less obscured type-2 QSOs (those
fit by IR22491) display colours akin to those of AGN-dominated or
"warm" SMGs. In the IRAC plot, the 8.0$/$4.5 ratio provides a good
measure of the AGN contribution (at z$\sim$3 the IRAC 8 $\mu$m band
corresponds to K rest-frame). Warm SMGs and less obscured type-2 QSOs
both display the highest 8.0$/$4.5 values as expected from hot-dust
contribution to the rest-frame K-band emission.  
In the IRAC$/$MIPS
plot (Fig. 3b) type-2 QSOs fit by IR22491 (less obscured) 
and ``warm'' SMGs show very similar colours.
Those type-2 QSOs that follow the SED of Arp220 (ie the
starburst component is more obscured) show the most extreme values in 24$/$8
and 5.8$/$3.6 with some type-2 QSOs falling outside the shaded area expected
for AGN type objects. We attribute the wide range variation in the
S$_{24}/$S$_{5.8}$ ratio to the
varying degree of obscuration, contribution from the central AGN and possibly
differences in dust composition. 

Sturm et al. (2006) presented mid-infrared IRS spectroscopy of a small
sample of primarily lower-redshift type-2 QSOs (0.2$<$z$<$1.3) but
also included CDFS-202.  They tried to estimate the AGN contribution
to the mid-infrared (6$\mu$m) part of the spectrum by performing a
decomposition of their IRS spectra using a starburst (M82) and and AGN
(linear continuum) template. Their results, together with the
surprisingly absence of PAH features in the IRS spectra, indicate that
the AGN contributes significantly (if not dominantly) in the
mid-infrared. More recently, Martinez-Sansigre et al (2008) presented
IRS spectra of a sample of z$\sim$2 type-2 QSOs (their selection was
based on radio and mid-infrared criteria, see Martinez-Sansigre et
al. 2006) and concluded that although they are predominantly continuum
dominated a large fraction shows deep silicate absorption features and
PAHs. The strength of the PAHs is similar to that found in
submillimetre luminous galaxies (e.g.  Valiante et al. 2007). The PAHs
are thought to originate in the host galaxy and are indicative of
vigorous star-forming activity. A similar conclusion has been reached
by Zakamska et al. (2008) based on mid-infrared spectroscopy of
optically selected type-2 QSOs. Combining our discussion here with the
SED fits we also conclude that the mid-infrared part of the spectrum
is dominated by emission from the AGN.

\begin{figure*}
\centering
\subfigure[] % caption for subfigure a
{
    \label{fig:sub:a}
    \includegraphics[width=60mm,angle=270]{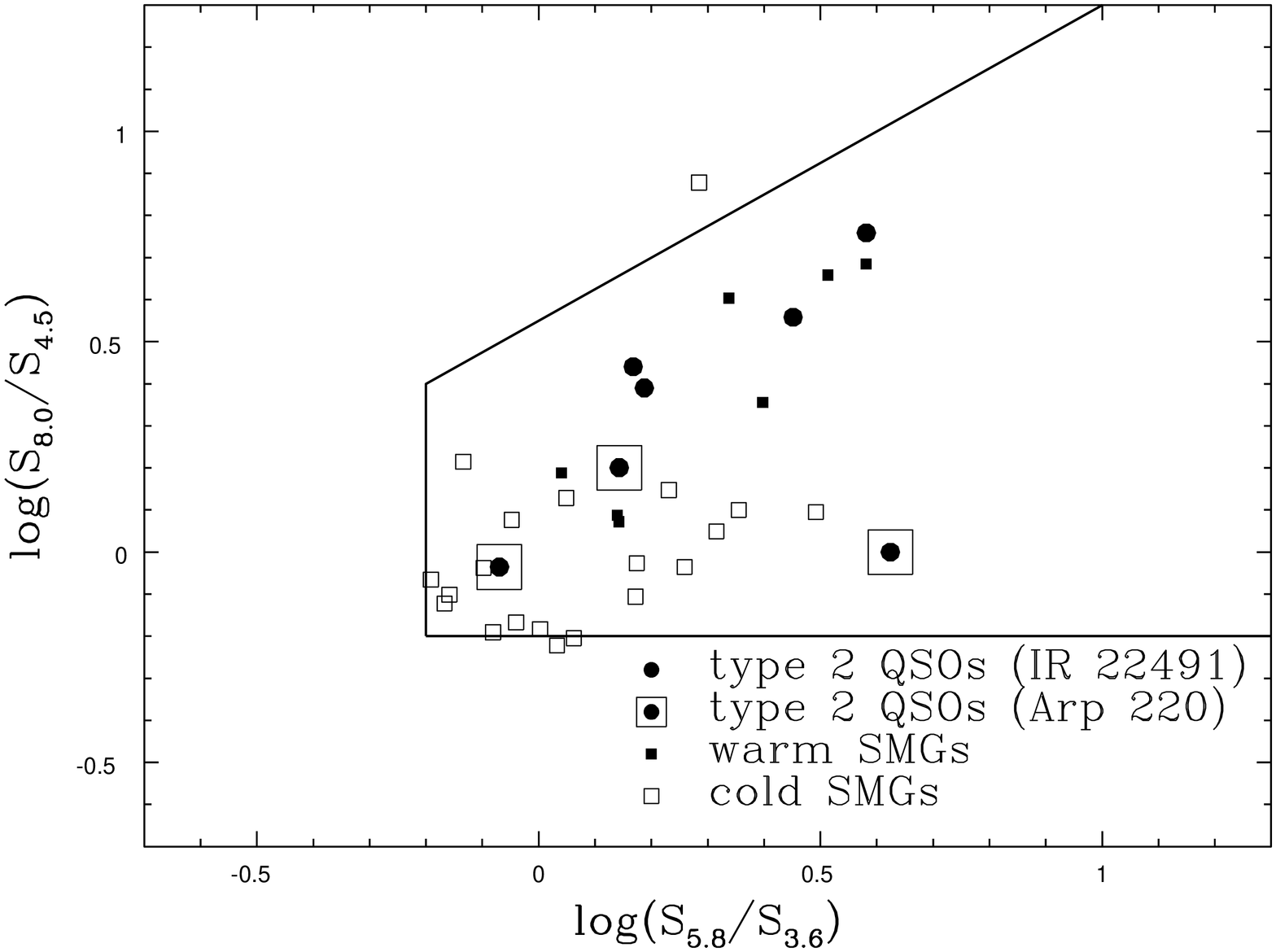}
}
\hspace{0.5cm}
\subfigure[] % caption for subfigure b
{
    \label{fig:sub:b}
    \includegraphics[width=60mm,angle=270]{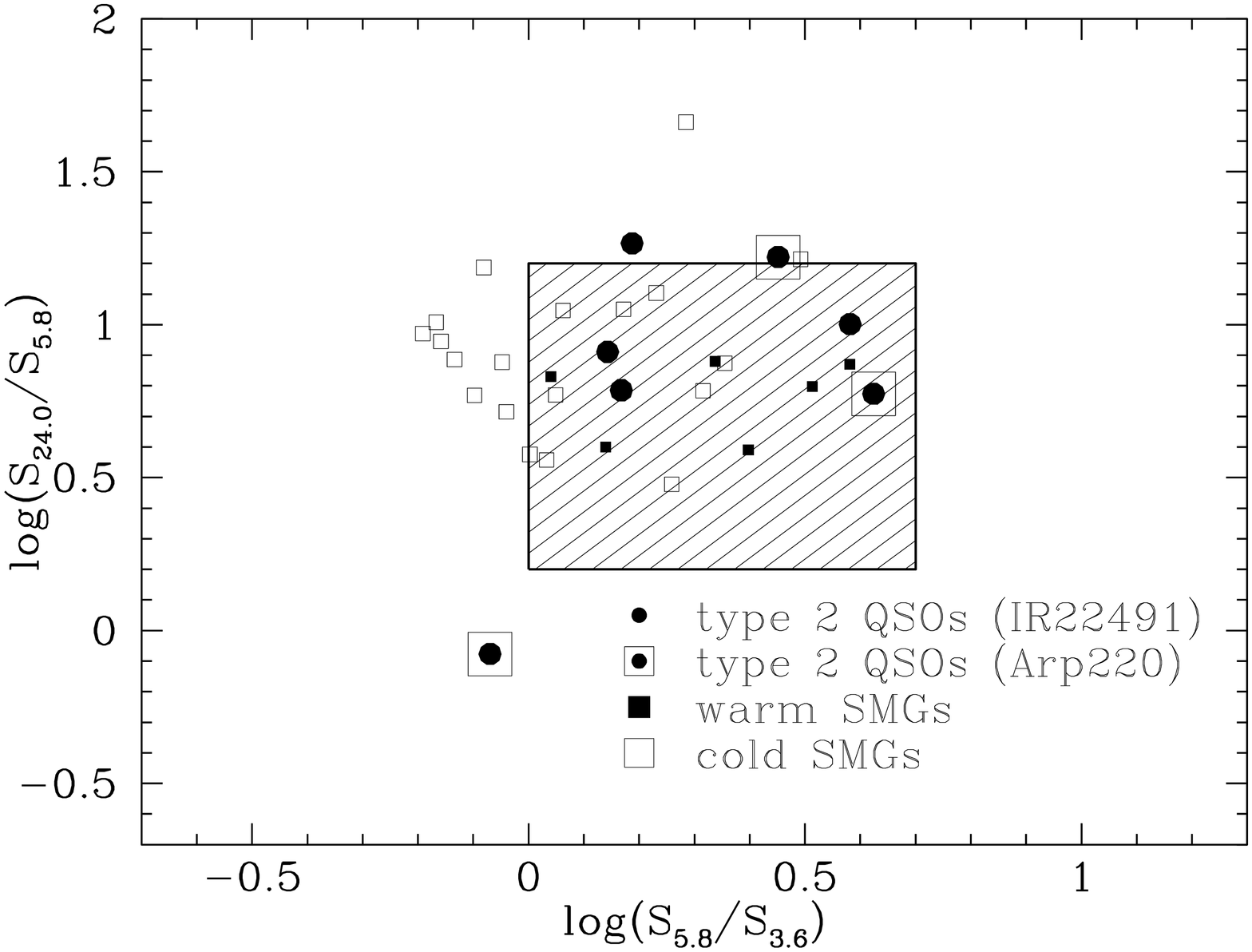}
}
\caption{Colour-colour plots involving IRAC (left) and IRAC and MIPS
bands (right, adapted from Lacy et al. 2004). Symbols are as follow:
filled circles: type-2 QSOs, filled squares: warm SCUBA sources, 
open squares: cold SCUBA sources (mid-infrared Spitzer data for SCUBA sources
from Egami et al. 2004 and Ashby et al. 2006). Those type 2 QSOs that follow
the Arp220 template SED (see discussion in section 3)
are marked with a squared circle. The wedged-shaped (plot 3a) and the 
hatched areas (plot 3b) denote the regions occupied by AGN.}
\label{fig:sub} %
\end{figure*}

\section{Far-Infrared Luminosities, Star Formation Rates and Masses}

In Table 3 we present far-infrared luminosities ($L_{FIR}$) for each source. 
We have used two different methods to estimate FIR luminosities based on:
converting the Spitzer 24 $\mu$m flux to ($L_{FIR}$) and, 
interpolating 
from the measured 850 $\mu$m flux density. 
Although L$_{FIR}$ estimates of  high-z objects carry 
uncertainties, depending on the method chosen, it is reassuring that the 
methods presently considered, yield within errors, similar results. 
In what follows we discuss each method in detail.

At the redshifts of our type-2 QSOs the Spitzer 24$\mu$m band corresponds
to rest-frame mid-IR (6--8 $\mu$m depending on redshift) which is, in general, 
used to infer the total thermal (8--1000 $\mu$m) 
$L_{FIR}$ (e.g. Chary \& Elbaz 2002).
A number of authors (e.g. Spinoglio et al. 1995, Rush et al. 1993)  have 
shown that, especially for AGN, 
the MIR (12$\mu$m) luminosity is a good proxy for the FIR luminosity. 
Following AH06 we estimate the 12$\mu$m-to-FIR luminosity conversion factor 
according to the two templates (Arp220 and IR22491) we used to fit the SEDs of
our type-2 QSOs. We derive 
the 12 $\mu$m rest-frame luminosity by extrapolating from the observed 
24 $\mu$m flux. 
The resulting $L_{FIR}$ values are listed in Table 3. All our type-2 QSOs 
are highly 
luminous, with luminosities formally in the ``ultra-luminous'' class 
($L_{FIR} > 10^{12}$L$_{\odot}$).

In section 3 we showed that the far-infrared$/$submm part of the SED
of type-2 QSOs can be well represented by the two ULIRG templates
Arp220 and IR22491. In both of these template galaxies submm emission
is emitted by dust heated by young stars (rather than the AGN
continuum). Under the assumption that the submm emission in type-2 QSOs is
also due to star-forming activity we can use the  850 $\mu$m fluxes
to estimate $L_{FIR}$ (by scaling them with those of
the templates used to fit each of the type-2 QSOs).  For IR22491 we use
log L$_{FIR}$=12.11 and for Arp220 log L$_{FIR}$=12.48 (e.g. Rigopoulou
et al. 1999). L$_{FIR}$ values estimated in this way are of course subjected to
uncertainties depending on the template of choice. We note however,
that the choice of template only affects the normalization of the
derived luminosities; its variation as a function of redshift over the
range covered by our sources (1.0$<$z$<$3.5) is small for both templates 
considered,
because the strong negative K-correction at 850 $\mu$m cancels out the
effect of cosmological dimming (e.g. Blain \& Longair 1993). 

Additionally, for those sources with radio detections we have used the 
well-established radio-far-infrared correlation to estimate $L_{FIR}$ with
values in good agreement with the previously discussed methods. 
We list the computed L$_{FIR}$ values in Table 3.

Using the L$_{FIR}$ values in Table 3 and following the prescription by 
Kennicutt (1998) we estimate SFR rates by 

$SFR = L_{FIR}  / (5.8 \times 10^{9} L_{\odot})$ ($M_{\odot}/yr$).

Using the observed 850 $\mu$m fluxes the dust mass is then given by:

$M_{dust} = (1+z)^{-1}S_{850}D_{L}^{2}/k_{rest}B(u_{rest, T_{d}})$

where z is the redshift, D$_{L}$ is the luminosity distance,
k$_{rest}$ is the rest frequency absorption coefficient and
B(v$_{rest, T_{d}}$ is the rest frequency value of the Planck function
from dust grains radiating at temperature T$_{d}$. Using
k$_{rest}$=0.15 (e.g. Scott et al. 2002) and assuming optically thin
thermal emission the derived M$_{dust}$ are listed in Table 3.

\begin{table*}
 \centering
 \begin{minipage}{160mm}
  \caption{FIR Luminosities, SFR, Dust and Black Hole Masses}
  \begin{tabular}{@{}cccccc@{}}
  \hline
ID &logL$_{FIR}$(12$\mu$m)$^{1}$&logL$_{FIR}$(850$\mu$m) &SFR$^{2}$&M$_{dust}$&
M$_{BH} ^{4}$\\
        &L$_{\odot}$ &L$_{\odot}$  &M$_{\odot}/yr$&$\times$10$^{8}$ M$_{\odot}$&
$\times$10$^{8}$ M$_{\odot}$   \\
\hline
54&12.58&12.78&670--1165 &4.9  &0.5\\
45&12.49 &12.72:$^{3}$ &540--905 &4.1 &1.4\\
263&12.55 &12.64 &620--1190 &4.2 &1.9\\
27&12.34 &12.62: &380--720 &5.1  &0.7\\
112&12.67 &12.88 &710--1370 &5.5 &0.8 \\
901&12.60 &12.87 &690--1460 &4.8 &0.6 \\
51&12.40: &12.79: &440--920 &3.9  &1.6\\
\hline
$^{1}$: luminosities based on 24 $\mu$m fluxes, see text\\
$^{2}$: range of SFR based on different L$_{FIR}$\\
$^{3}$: (:) denotes upper limit\\
$^{4}$: black hole masses are lower limits
\end{tabular}
\end{minipage}
\end{table*}

Finally, following Marconi et al. (2004) we estimate bolometric
luminosities based on rest-frame absorption corrected
L$_{X}$. Assuming that L$_{bol}$ = L$_{Edd}$, where L$_{Edd}$ is the
Eddington luminosity we report (Table 3) lower limits on the black
hole masses.

\section{Origin of the Submillimetre Emission}

A different approach to the relative importance of AGN and star formation
activity is provided by comparing the
submillimetre (850 $\mu$m) to X-ray (2 keV) spectral slope
($\alpha_{SX}$) of our type-2 QSOs to those of template nearby galaxies. 
We have, therefore, calculated the $\alpha_{SX}$ values for our sample 
galaxies using:
\\
%\begin{equation}
$
\alpha_{SX} = - log [\frac{f_{2KeV}}{f_{850}} - 0.18] \times 0.163
$
%\end{equation}
\\
where f$_{2keV}$ is the observed flux density at 2 keV 
(keV cm$^{-2}$ s$^{-1}$ keV$^{-1}$), calculated from the full-band flux 
using the estimated photon index, and f$_{850}$ is the observed flux density 
at 850 $\mu$m (mJy). 
Figure 4 shows the estimated $\alpha_{SX}$ values as a function of
redshift for our sample type-2 QSOs. In the same plot we also show 
template SEDs for: an
unabsorbed quasar, quasars with varying degrees of absorption (N$_{H}$
= 10$^{23}$, 10$^{24}$ cm$^{-2}$ and Compton-thick with 1\% of the
nuclear emission scattered), a mean starburst and an
Arp220-like. 
\begin{figure*}
\includegraphics[width=80mm,angle=270]{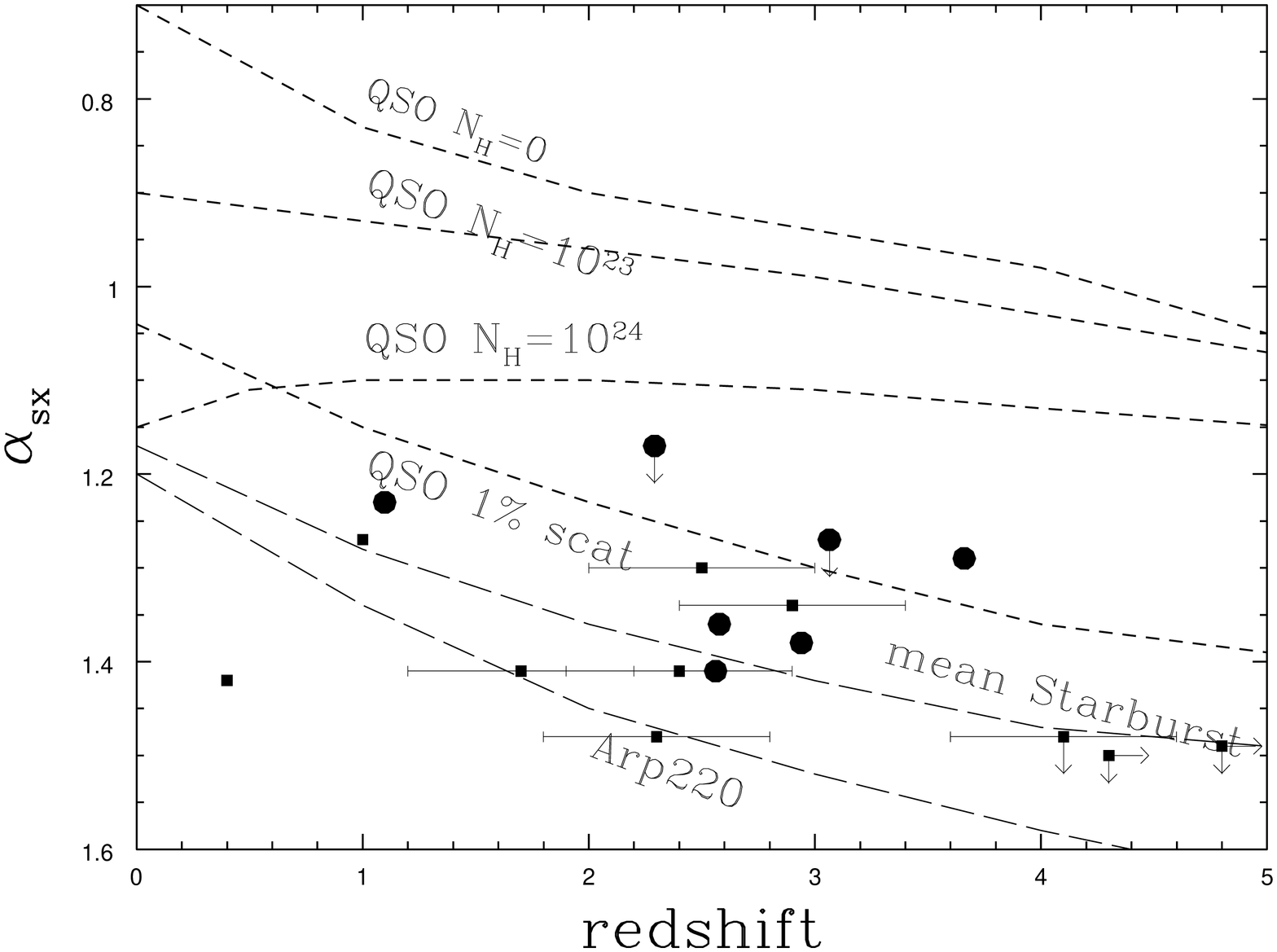}
 \caption{ Submillimetre to X-ray spectral index ($\alpha_{SX}$ as a function
 of
redshift. The curves show the expected $\alpha_{SX}$ values for a set of SEDs
(QSOs with various degrees of obscuration, mean Starburst, Arp220). 
The values
for the present type 2 QSOs are denoted by solid squares. Filled circles
correspond to SCUBA galaxies from Alexander et al. (2005).}
\end{figure*}

The templates used here are discussed in more details in Almaini et
al. (2003). The type-2 QSOs yield values ranging between 1.41 (for
CDFS-901) to 1.17 (for CDFS-45).  Larger values of $\alpha_{SX}$ indicate
stronger submillimetre emission relative to the X-ray emission. 
The
$\alpha_{SX}$ indices are clearly incompatible with unabsorbed AGN.  The
$\alpha_{SX}$ index for CDFS901, CDFS54, CDFS51 and CDFS112 are compatible
with the mean starburst and Arp220 templates. It is of interest to
note that of these four sources three are best fit by the Arp220
template (see discussion in section 3). The remaining sources CDFS263,
CDFS 27 and CDFS45, have indices that are close to the spectral index
of a Compton-thick AGN in which only 1\% of the nuclear emission is
seen through scattering ( the properties of CDFS263 have already been
discussed in Mainieri et al. 2005). 
For comparison we report in Figure 4 the spectral indices calculated for the
submillimetre sources studied by Alexander et al. (2005). These
sources show a wide range in indices similar to what we reported for
our type 2 QSOs. In terms of absolute $\alpha_{SX}$ value, the four
type-2 QSOs we discussed before are indistinguishable from the
submillimetre sources. 

Another approach to assess the relative importance of the two
components (AGN and starburst) is to examine the x-ray-to-far-infrared
luminosity ratio.  While this ratio is in principle quite similar to
the $\alpha_{SX}$ spectral index, it is independent of the X-ray
spectral slope (which one has to assume when converting broad band to
monochromatic flux densities). In Figure 5 we plot the $L_{x}$ vs
$L_{FIR}$ ratio for the present sample of type-2 QSOS (with values
taken from Tables 2 and 3) and literature galaxies which also host
obscured AGN. Such a comparison is instructive with the caveat that
the $L_{FIR}$ estimates may originate from a variety of measurements
(e.g mid-infrared, submm, etc). In the same plot we show a mean
''starburst'' $L_{x}/L_{FIR}$ ratio (adapted from the work of David,
Jones and Forman 1992) and the same ratio for quasars (from Elvis et
al. 1994). It is difficult to define the same ``mean'' $L_{x}/L_{FIR}$
ratio for AGN since they show such a huge dispersion in their
values. Additionally, we have marked the location of Arp220, Mrk231
and NGC 6240. The submillimetre sources (in particular those whose
X-rays originate from an AGN, Alexander et al.  2005) display values
very similar to those of the present type-2 QSOs.  For further
comparison we have also marked the location of the SDSS optically
selected type-2 QSOs (Zakamska et al. 2008). These show a wide spread
in their $L_{x}/L_{FIR}$, spanning from values as low as those
displayed by starbursts all the way to the quasar values.  None of the
type-2 QSOs have $L_{x}/L_{FIR}$ values larger than those of the
quasars presented in Elvis et al. (1994). This finding implies that
type-2 QSOs are likely to experience some level of
 star-formation activity which
could be responsible for the $FIR$ and submillimetre emission.  In
turn, such star-forming processes could well contribute
to the total bolometric output although, at this stage we
cannot accurately attribute a specific fraction. Within our very small
sample we do, however, see a range in the strength of the starburst
which may reflect the evolutionary stage of each object. In the next
section we investigate in more detail a possible evolutionary scheme for these
sources. 

\begin{figure*}
\includegraphics[width=80mm,angle=270]{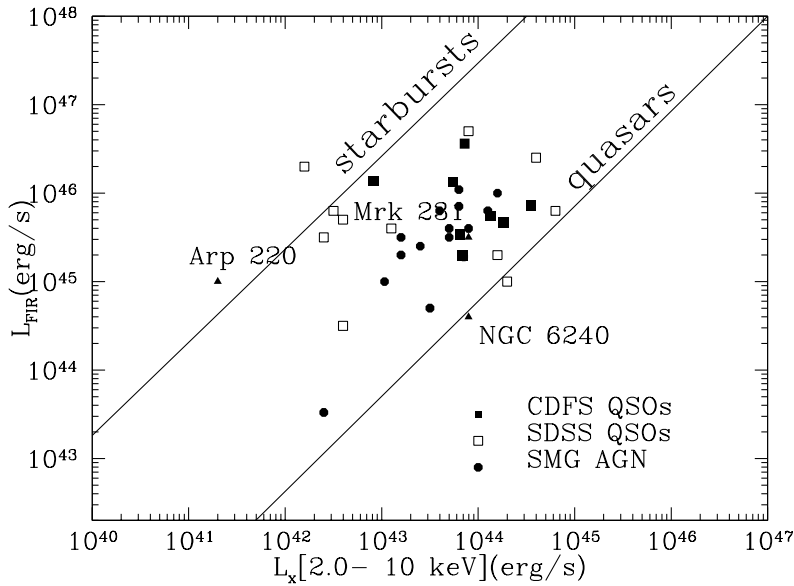}
 \caption{ Rest-frame far-infrared vs. absorption-corrected [2-10] 
keV luminosity for
our type-2 QSOs (filled squares) and other obscured AGN from the literature. 
SDSS optically selected type-2 QSOs (open squares) from Zakamska et al. (2008),
Submillimetre galaxies (filled circles) from Alexander et al. (2005), local
Universe ``template'' objects, Arp 220, Mrk 231 and NGC 6240 
(filled triangles). The diagonal lines show ratios of X-ray to far-infrared
luminosity for Starbursts (adapted from the work of David, Jones and 
Forman 1992, the X-ray data have been converted to the [2--10] band used here 
assuming
$\Gamma$ = 2.0) and quasars (adapted from Elvis et al. 1994).
}
\end{figure*}

\section{The evolution of type-2 QSOs}

It has been suggested that the early phases of QSO evolution are
characterised by substantial absorption (e.g. Fabian 1999). In such
models the main obscured growth phase of the QSO coincides with the
formation of the galaxy spheroid. Once this phase is completed then
the QSO blows away the obscuring material and starts its luminous
unobscured phase (e.g. Silk \& Rees 1998). Such models deviate from
the traditional view of the ``unified scheme'' for local AGN (e.g.
Antonucci et al. 1993) whereby the absorbing material is present in 
all objects in the form of a toroidal structure. Based on simulations
of galaxy mergers that include black holes (e.g. di Matteo et al. 2005; 
Springel et al. 2005), Hopkins et al. (2005) determine the intrinsic
lifetime of the strong accretion phase and, taking into account obscuration,
infer a lifetime for the QSO. They find that the optical QSO phase is rather 
short compared to the intrinsic lifetime of the quasar. Additionally, 
their model suggests the presence of an obscured phase (type 2 QSOs) that
occurs immediately before the dust is completely blown away.

By contrasting the X-ray$/$submillimetre properties of an X-ray absorbed, 
Compton-thin QSO sample with that of a matched unabsorbed sample of QSOs,
Page et al. (2004) and Stevens et al. (2005) provided observational
evidence for the last two stages
of the evolutionary scenario described above. The X-ray absorbed QSOs 
are observed during the transition phase, where the QSO is still actively
forming stars (formation of the galactic bulge) while at the same time
accretes material that assists the formation of the central black hole.
Once the black hole has reached a mass 
$\geq$ few $\times$ 10$^{8}$M$_{\odot}$
the QSO enters its unobsucred luminous phase, slowly switching off the
star-forming process and eventually evolving into relaxed elliptical galaxies
(e.g. Kukula et al. 2001, Dunlop et al. 2003).

With bolometric luminosities in excess of $L_{bol} > 10^{46} erg s^{-1}$, 
obscuring column densities N$_{H} >$10$^{23}$cm$^{-2}$, the
present type-2 QSO sample can be used to further investigate the QSO
evolutionary scheme. In section 3 we compared the 
SEDs of type-2
QSOs with those of local ULIRG templates IR22491
and Arp220. Local ULIRGs are characterised by extremely high
bolometric luminosities and similarly large column densities.  If we
convert the estimated A$_{V}$ of ULIRGs (e.g. from Genzel et al. 1998) to
expected N$_{H}$ that results in N$_{H}$ values in excess of
10$^{23}$cm$^{-2}$.  
Given the high obscuration, the present (albeit limited at the moment) 
detections in the submillimetre, large 
dust masses (and consequently gas) and high redshifts, the present 
sample of type-2 QSOs could be candidates
for the initial highly obscured phase of QSO evolution. A
pre-requisite for any QSO to be ``caught'' in this phase is of course
submillimetre emission.  During this phase the black hole must be
accreting rapidly in order to achieve a mass of
$\sim$10$^{8}$M$_{\odot}$ and thus should also reach X-ray
luminosities in excess of 10$^{44}$ ergs$^{-1}$.  And likewise the
local ULIRGs, the properties of QSOs in this initial phase of
evolution are not homogeneous in terms of luminosity, submillimetre
brightness, even the starburst$/$AGN contribution.

Is there a link between type-2 QSOs and submm-bright galaxies?  In
section 4 we compared the mid-infrared properties of the two samples
and found that they show certain similarities. Additional evidence was
provided in Figures 4 and 5 through the comparison of the $\alpha_{SX}$ indices
and the x-ray-to-far-infrared luminosity. For
more than half of the type-2 QSOs their $\alpha_{SX}$ index is identical to
that measured for submillimetre-bright objects from the Alexander et
al. (2005) sample. It is thus quite possible that a fraction of the
submm-luminous sources are in fact caught during this initial highly
absorbed and actively star-forming phase in the QSO evolution. The
evidence provided by Alexander et al. (2005) on the existence of black
holes in a fraction of the submillimetre-bright population gives
further support to this claim. Likewise, we have found that the 
L$_{x}/$L$_{FIR}$ ratio of the present type-2 QSOs is smaller than that
found for the well studied quasars of Elvis et al. (or objects that
are purely AGN dominated).

We have presented some evidence for the initial stages of QSO
evolution by reporting measurements of submillimetre emission from
a sample of highly absorbed type-2 QSOs discovered through deep X-ray 
surveys. Two out of the eight targets have firm detections ($S/N>$4) and a 
further three objects are marginally detected.
These submillimetre detections coupled with their overall
similarities to local ULIRGs suggest that these type-2 QSOs
may be experiencing co-eval growth of their spheroid component while at the
same time maintaining an accretion rate high enough to form a black
hole of M$_{BH}\sim$10$^{8}$M$_{\odot}$. Once this is achieved, the QSO
will blow away the obscuring material revealing its luminous QSO and
following a passive evolution into the present-day ellipticals.
Future sensitive submillimetre observations are clearly needed to further 
strengthen this claim. Deep sensitive surveys at 250, 320 and 520 $\mu$m with
HERSCHEL and SCUBA -2 (450 and 850 $\mu$m) will be crucial in delineating 
the shape of the far-infrared$/$submm part of the SED of type-2 QSOs and 
provide firm evidence of their submillimetre emission.

%\section{Conclusions}

\section*{Acknowledgments}
We thank the anonymous referee for useful comments on an earlier
version of this work. We thank Ben Weiner for useful discussions on the 
draft. This research has made use of the NASA/IPAC
Extragalactic Database (NED), which is operated by the Jet Propulsion
Laboratory, California Institute of Technology, under contract with
the National Aeronautics and Space Administration.

{}

\end{document}